\documentclass[english,pra,aps,amsmath,amssymb,showpacs,preprint]{revtex4}
\usepackage[T1]{fontenc}
\usepackage[latin9]{inputenc}
\usepackage{amsmath}
\usepackage{graphicx}
\usepackage{amssymb}
\usepackage{esint}

\makeatletter
\@ifundefined{textcolor}{}
{%
 \definecolor{BLACK}{gray}{0}
 \definecolor{WHITE}{gray}{1}
 \definecolor{RED}{rgb}{1,0,0}
 \definecolor{GREEN}{rgb}{0,1,0}
 \definecolor{BLUE}{rgb}{0,0,1}
 \definecolor{CYAN}{cmyk}{1,0,0,0}
 \definecolor{MAGENTA}{cmyk}{0,1,0,0}
 \definecolor{YELLOW}{cmyk}{0,0,1,0}
 }

\usepackage{subfigure}\usepackage{hyperref}\usepackage{float}\hypersetup{
unicode=false,
pdftoolbar=true,
pdfmenubar=true,
pdffitwindow=false,
pdfstartview={FitH},
pdftitle={Mytitle},
pdfauthor={Author},
pdfsubject={Subject},
pdfcreator={Creator},
pdfproducer={Producer},
pdfkeywords={keywords},
pdfnewwindow=true,
colorlinks=true,
linkcolor=red,
citecolor=blue,
filecolor=magenta,
urlcolor=cyan
}

\makeatother

\usepackage{babel}

\makeatother

\usepackage{babel}

\makeatother

\usepackage{babel}

\makeatother

\usepackage{babel}

\makeatother

\usepackage{babel}

\begin{document}

\title{Geometric phases for nonlinear coherent and squeezed states}

\author{Da-Bao Yang }

\affiliation{Theoretical Physics Division, Chern Institute of Mathematics, Nankai
University, Tianjin 300071, People's Republic of China}

\author{Ying Chen}

\affiliation{Theoretical Physics Division, Chern Institute of Mathematics, Nankai
University, Tianjin 300071, People's Republic of China}

\author{Fu-Lin Zhang}

\affiliation{Physics Department, School of Science, Tianjin University, Tianjin
300072, People's Republic of China}

\author{Jing-Ling Chen}

\email{chenjl@nankai.edu.cn}

\affiliation{Theoretical Physics Division, Chern Institute of Mathematics, Nankai
University, Tianjin 300071, People's Republic of China}

\date{\today}
\begin{abstract}
The geometric phases for standard coherent states which are widely
used in quantum optics have attracted a large amount of attention.
Nevertheless, few physicists consider about the counterparts of non-linear
coherent states, which are useful in the description of the motion
of a trapped ion. In this paper, the non-unitary and non-cyclic geometric
phases for two nonlinear coherent and one squeezed states are formulated
respectively. Moreover, some of their common properties are discussed
respectively, such as gauge invariance, non-locality and non-linear
effects. The non-linear functions have dramatic impacts on the evolution
of the corresponding geometric phases. They speed the evolution up
or down. So this property may have application in controlling or measuring
geometric phase. For the squeezed case, when the squeezed parameter
$r\rightarrow\infty$, the limiting value of the geometric phase is
also determined by non-linear function at a given time and angular
velocity. In addition, the geometric phases for standard coherent
and squeezed states are obtained under a particular condition. When
the time evolution undergoes a period, their corresponding cyclic
geometric phases are achieved as well. And the distinction between
the geometric phases of the two coherent states maybe regarded as
a geometric criterion. 
\end{abstract}


\maketitle

\section{Introduction}

\label{sec:introduction}

Recently, nonlinear coherent state and squeezed state have drew lots
of attentions of physicists. Filho and Vogel \cite{filho1996nonlinear}
defined the nonlinear coherent states as the right eigenstate of a
generalized annihilation, whose concrete definition will be given
in the following paragraph. They possessed not only some of the typical
features of the standard coherent states but also the strong non-classical
properties. B. Roy and P. Roy \cite{roy2000new} used displaced-operator
technique to discover a new nonlinear coherent sates. Its non-classical
properties, such as quadrature squeezing, amplitude-squared squeezing
and sub-Poissonian behavior, was studied intensively. Subsequently,
Kwek and Kiang \cite{kwek2003squeezed} researched its nonlinear extension
of the single-mode squeezed states and showed its sub-Poissonian behavior.

The above paragraph reviewed the developments of the nonlinear coherent
and squeezed state, now we pay attention to the evolution of geometric
phases. In retrospect, it is discovered by Berry \cite{berry1984quantal}
in the context of adiabatic, unitary, cyclic evolution of time-dependent
quantum system. Besides of the significance of physics, it also has
a geometrical interpretation which is given by Simon \cite{simon1983holonomy}.
It can be regarded as the holonomy of a line bundle $L$ over the
space of parameters $M$ of the system, if $L$ is endowed with a
natural connection.

Berry's result was generalized by Aharonov and Anandan \cite{aharonov1987phase}
via discarding the assumption of adiabaticity. The Aharonov and Anandan
phase (A-A phase) could be obtained by the difference between the
total phase and the dynamical one as well as be determined by the
loop integral of a natural connection on a $U(1)$ principle fiber
bundle over projective Hilbert space.

Afterwards, A-A phase was also generalized by Samuel and Bhandari
\cite{samuel1988general}. Specifically speaking, depending on Pancharatnam's
earlier work \cite{pancharatnam1956connection}, they found a more
general phase in the context of non-cyclic and non-unitary quantal
evolution. However, their definition is an indirect one, which depends
on a geodesic closing the initial and the final points of the open
path. Moreover, it is not manifestly gauge invariant. Soon Pati \cite{pati1995gauge,pati1995geometric}
gave another definition without the need of geodesic, where a canonical
one-form was defined and its line integral gave the geometric phase.
It is manifestly invariant under phase and gauge transformation.

However, when the initial and final states are orthogonal, the above
non-cyclic geometric phases become undefined. Manini and Pistolesi
\cite{manini2000offdiagonal} introduced abelian off-diagonal geometric
phases under the condition of adiabatic evolution. Mukunda et. al.
\cite{mukunda2001bargmann} generalized the concepts without the confinement
of adiabaticity. Furthermore, Kult et. al. \cite{kult2007nonabelian}extended
the conceptions to the non-abelian cases.

Another line of development of geometric phase is the generalization
to mixed states. Uhlmann \cite{uhlmann1986parallel} discussed this
issue in the mathematical context of purification. Sj$\ddot{o}$qvist
et. al. \cite{sjoqvist2000geometric} addressed the non-degenerate
geometric phase in non-cyclic and unitary evolution under the background
of quantum interference. Some other extensions \cite{tong2003geometric,tong2004kinematic}
were also made.

Including applications of the geometric phases in many areas of physics
\cite{shapere1989geometric,bohm2003geometric}, new perspectives such
as topological quantum computing \cite{nayak2008nonabelian} and phase
control \cite{abdel2009phase,abdel2010sensitivity} were opened as
well.

Based on the formulation given by Pati \cite{pati1995gauge,pati1995geometric},
the present paper aims at giving non-cyclic and non-unitary geometric
phases of nonlinear coherent and squeezed states. It is organized
as follows. Sec. II reviews nonlinear coherent and squeezed states
as well as the non-unitary and non-cyclic geometric phase. In Sec.
III, the geometric phases of the two nonlinear coherent states are
calculated respectively. Moreover we analyze the distinction between
the two geometric phase and some of their properties. In Sec. IV,
the geometric phase of squeezed state is given. In addition, some
of its features are discussed. At the end of this paper, a conclusion
is drawn.

\section{reviews of nonlinear coherent and squeezed states and geometric phases}

\label{sec:reviews}

For nonlinear coherent states, the generalized annihilation and creation
operators are introduced by\[
A=af(N),~~~~~A^{\dagger}=f(N)a^{\dagger},\]
 where $a(a^{\dagger})$ is the harmonic oscillator annihilation (creation)
operator, $N=a^{\dagger}a$, and $f(x)$ is a reasonably well behaved
real function. $A$ and $A^{\dagger}$ satisfy the following nonlinear
algebra\begin{equation}
[A,~A^{\dagger}]=(N+1)f^{2}(N+1)-Nf^{2}(N).\label{eq:Commutator}\end{equation}
 So we can construct the coherent operator, i.e., $\exp(\alpha A^{\dagger}-\alpha^{*}A).$
However, because of Eq. \eqref{eq:Commutator}, we can't apply BCH
disentangling theorem to obtain the coherent operator \cite{roy2000new}.
So we construct another two operators $B$ and $B^{\dagger}$ which
have the following commutation relations\[
[A,~B^{\dagger}]=1\]
 and\[
[B,~A^{\dagger}]=1,\]
 where $B=a\frac{1}{f(N)}$ and $B^{\dagger}=\frac{1}{f(N)}a^{\dagger}$.
As a result, the coherent state operators \cite{roy2000new} can be
constructed as\[
D_{1}(\beta)=\exp(\beta B^{\dagger}-\beta^{*}A)\]
 and\[
D(\beta)=\exp(\beta A^{\dagger}-\beta^{*}B).\]
 It is worthy of note that the above two operators are not unitary.
When both of them act on the vacuum state, we can get two coherent
states which are\[
|\beta_{1}\rangle=D_{1}(\beta)|0\rangle=\exp(\beta B^{\dagger}-\beta^{*}A)|0\rangle=\exp\left(-\frac{|\beta|^{2}}{2}\right)\exp(\beta B^{\dagger})\exp(-\beta^{*}A)|0\rangle\]
 and\[
|\beta\rangle=D(\beta)|0\rangle=\exp(\beta A^{\dagger}-\beta^{*}B)|0\rangle=\exp\left(-\frac{|\beta|^{2}}{2}\right)\exp(\beta A^{\dagger})\exp(-\beta^{*}B)|0\rangle\]
 via BCH theorem. Both of them are represented in the energy eigenstates,
i.e.,\begin{equation}
|\beta_{1}\rangle=e^{-\frac{|\beta|^{2}}{2}}\sum_{n=0}^{\infty}\beta^{n}\frac{1}{f(n)!\sqrt{n!}}|n\rangle\label{eq:FirstCoherentState}\end{equation}
 and\begin{equation}
|\beta\rangle=e^{-\frac{|\beta|^{2}}{2}}\sum_{n=0}^{\infty}\beta^{n}\frac{f(n)!}{\sqrt{n!}}|n\rangle\label{eq:SecondCoherentState}\end{equation}
 where $f(n)!=f(1)\cdots f(n).$ If $f(n)$ takes a particular form,
the first nonlinear coherent state $|\beta_{1}\rangle$ is regarded
as stationary states of a center of mass motion of a trapped and bi-chromatically
laser-driven ion far from the so-called Lamb-Dicke regime. By comparison
with the standard coherent states, the nonlinear counterparts exhibit
non-classical properties such as strong amplitude squeezing and self-splitting
into two or more sub-states, which finally give birth to quantum interferences
\cite{filho1996nonlinear}. With respect to the second nonlinear coherent
state $|\beta\rangle$, it still displays various non-classical properties
such as quadrature squeezing, amplitude-squared squeezing and sub-Poissonian
behaviour \cite{roy2000new}.

Furthermore, Kwek and Kiang \cite{kwek2003squeezed} pointed out that
the squeezed operators could be defined as\[
S_{1}(\zeta)=\exp\left[\frac{1}{2}(\zeta B^{\dagger2}-\zeta^{*}A^{2})\right]\]
 and\[
S(\zeta)=\exp\left[\frac{1}{2}(\zeta A^{\dagger2}-\zeta^{*}B^{2})\right].\]
 By the same procedure, when the above two operators act on the vacuum
state, one can get the so-called squeezed states, i.e.,\[
|\zeta_{1}\rangle=S_{1}(\zeta)|0\rangle=\exp\left[\frac{1}{2}(\zeta B^{\dagger2}-\zeta^{*}A^{2})\right]|0\rangle\]
 and\[
|\zeta\rangle=S(\zeta)|0\rangle=\exp\left[\frac{1}{2}(\zeta A^{\dagger2}-\zeta^{*}B^{2})\right]|0\rangle.\]
 If the above two states are expanded in the energy eigenstates, we
can find that they coincide with each other, that is to say,\[
|\zeta\rangle=|\zeta_{1}\rangle=\mathcal{N}\sum_{n=0}^{\infty}\left(\frac{e^{i\varphi}\tanh(r)}{2}\right)^{n}\frac{f(2n)!\sqrt{(2n)!}}{n!}|2n\rangle,\]
 where $\mathcal{N}$ is an indefinite coefficient and $\zeta=re^{i\varphi}$.
By the way, the following formulas maybe useful in expanding the squeezed
states, which are\[
\exp\left[\frac{1}{2}(\zeta B^{\dagger2}-\zeta^{*}A^{2})\right]A\exp\left[-\frac{1}{2}(\zeta B^{\dagger2}-\zeta^{*}A^{2})\right]=\cosh(r)A-e^{i\varphi}\sinh(r)B^{\dagger}\]
 and\[
\exp\left[\frac{1}{2}(\zeta A^{\dagger2}-\zeta^{*}B^{2})\right]B\exp\left[-\frac{1}{2}(\zeta A^{\dagger2}-\zeta^{*}B^{2})\right]=\cosh(r)B-e^{i\varphi}\sinh(r)A^{\dagger}.\]
 Moreover, it is pointed out that unlike the normal squeezed states,
the Mandel Q number of the non-linear case is negative and as a consequence
it exhibits sub-Poissonian statistics.

The above paragraphs are about nonlinear coherent and squeezed states.
And it is well-known that the operators are non-unitary. Hence, to
calculate the geometric phases, we must know the methods of calculating
the non-unitary case, which are reviewed in the following paragraphs.
Moreover, it is also applied to non-cyclic evolution.

For a non-cyclic evolution, a way must be found to compare two different
quantal state. According to Pancharatnam connection \cite{pancharatnam1956connection},
the total phase \cite{pati1995gauge} can be formulated as \begin{equation}
\chi=\arg\left\langle \frac{\psi(0)}{\parallel\psi(0)\parallel}\mid\frac{\psi(t)}{\parallel\psi(t)\parallel}\right\rangle ,\label{eq:TotalPhase}\end{equation}
 where $\parallel\psi(t)\parallel$ denotes the norm of the state
vector $|\psi(t)\rangle$ and $|\psi(t)/\left\Vert \psi(t)\right\Vert \rangle$
represents $|\psi(t)\rangle/\left\Vert \psi(t)\right\Vert $. And
the dynamical phase can be written as\[
\delta=-i\intop_{\tau=0}^{t}\left\langle \frac{\psi(\tau)}{\parallel\psi(\tau)\parallel}|\frac{d}{d\tau}|\frac{\psi(\tau)}{\parallel\psi(\tau)\parallel}\right\rangle d\tau.\]
 Hence the geometric phase \cite{mukunda1993quantum} reads\begin{equation}
\gamma=\chi-\delta.\label{eq:PreviousGeometricPhase}\end{equation}
 In order to spell out the geometric aspects of the above equation,
Pati \cite{pati1995gauge,pati1995geometric} introduce the reference
section which is\begin{equation}
|\chi_{0}(t)\rangle=\frac{\langle\psi(t)|\psi(0)\rangle}{|\langle\psi(t)|\psi(0)\rangle|}\frac{|\psi(t)\rangle}{\parallel\psi(t)\parallel}.\label{eq:ReferenceSection}\end{equation}
In order to understand the above definition, we give a brief outline
of fiber bundles. A quantum state $\rho$ and a relative phase are
represented by a point of projective Hilbert space $\mathcal{P}$
and a circle $U(1)$ respectively. Take the Cartesian product of $\mathcal{P}$
and $U(1)$ as well as define a projection map $\Pi:\mathcal{L}\rightarrow\mathcal{P}$,
where $\mathcal{L}=\mathcal{P}\times U(1)$. Hence, we can regard
the triple $\mathcal{L}(\mathcal{P},U(1),\Pi)$ as a fiber bundle,
in which $\mathcal{P}$ and $U(1)$ are termed the base space and
the fiber respectively. Furthermore, the evolution of quantum state
$\rho(t)$ can generate a curve in $\mathcal{P}$. Then, there exists
another map $s:\mathcal{P}\rightarrow\mathcal{L}$ such that the image
$\Gamma(t)$ of curve $\rho(t)\in\mathcal{P}$ lies in the fiber over
$\rho$, i.e., $\Pi\circ s=Id_{\mathcal{P}}$. And the map is called
section. It is easy to know $|\chi_{0}(t)\rangle$ is a section of
the fiber bundle $\mathcal{L}$. If $|\chi_{0}(t)\rangle$ is going
to be defined as reference section, it has to satisfy the following
condition that $\langle\chi_{0}(0)|\chi_{0}(t)\rangle$ is always
real and positive during the evolution, in another word the two state
vectors are in phase, which is in accordance with the Pancharatnam
condition. Moreover, by use of Eq. \eqref{eq:ReferenceSection}, the
non-cyclic and non-unitary geometric phase \cite{pati1995geometric}
could be expressed as\begin{equation}
\gamma=i\intop_{\tau=0}^{t}\left\langle \chi_{0}(\tau)|\frac{d}{d\tau}|\chi_{0}(\tau)\right\rangle d\tau~mod~2\pi.\label{eq:GeometricPhase}\end{equation}
It is not very complex to verify that the geometric phase is independent
of the choice of the phase of the quantum system state $|\psi(t)\rangle$,
reparametrization invariant and manifestly gauge invariant. Furthermore
the geometric phase could reduce to Berry phase and A-A phase under
a appropriate limit.

Up to now, we have already reviewed both the nonlinear coherent and
squeezed state, as well as non-cyclic and non-unitary geometric phase.
Hence we can research further on the geometric phases of nonlinear
coherent and squeezed states.

\section{geometric phase for nonlinear coherent state}

\label{sec:conherent}

At any time $t>0$, the first coherent state vector is given by\begin{equation}
|\beta_{1}(t)\rangle=e^{-\frac{|\beta|^{2}}{2}}\sum_{n=0}^{\infty}\beta^{n}\frac{1}{f(n)!\sqrt{n!}}e^{-iE_{n}t}|n\rangle,\label{eq:TimeEvolutionOfFirstCoherentState}\end{equation}
 where $E_{n}=\omega(\frac{1}{2}+n)$ and we set $\hbar=1$ for simplicity.
In order to calculate the corresponding geometric phase, we must know
the reference section \eqref{eq:ReferenceSection}. At first, let's
calculate the first fraction of Eq. \eqref{eq:ReferenceSection}.
Considered about this problem in a polar coordinate, the first fraction
is expressed as\begin{equation}
\frac{\langle\beta_{1}(t)|\beta_{1}(0)\rangle}{|\langle\beta_{1}(t)|\beta_{1}(0)\rangle|}=e^{-i\chi}=\exp\left[i\arctan(\frac{Im(\langle\beta_{1}(t)|\beta_{1}(0)\rangle)}{Re(\langle\beta_{1}(t)|\beta_{1}(0)\rangle)})\right],\label{eq:FirstFraction}\end{equation}
 Where $\chi$ is the total phase in accordance with Eq. \eqref{eq:TotalPhase}.
So it is necessary to obtain the inner product \begin{equation}
\langle\beta_{1}(t)|\beta_{1}(0)\rangle=e^{-|\beta|^{2}}\sum_{n=0}^{\infty}\frac{|\beta|^{2n}}{[f(n)!]^{2}n!}\cos(E_{n}t)+ie^{-|\beta|^{2}}\sum_{n=0}^{\infty}\frac{|\beta|^{2n}}{[f(n)!]^{2}n!}\sin(E_{n}t).\label{eq:InnerProduct}\end{equation}
 Substituting Eq. \eqref{eq:InnerProduct} into Eq. \eqref{eq:FirstFraction},
one can gets\begin{equation}
-\chi=\arctan\frac{\sum_{n=0}^{\infty}\frac{|\beta|^{2n}}{[f(n)!]^{2}n!}\sin(E_{n}t)}{\sum_{n=0}^{\infty}\frac{|\beta|^{2n}}{[f(n)!]^{2}n!}\cos(E_{n}t)}.\label{eq:MinusTotalPhase}\end{equation}
 Second, to seek for $\parallel\beta_{1}(t)\parallel$, we can acquire
it in this way, i.e.,\begin{equation}
N_{1}=\parallel\beta_{1}(t)\parallel=\sqrt{\langle\beta_{1}(t)|\beta_{1}(t)\rangle}=e^{-\frac{|\beta|^{2}}{2}}\left\{ \sum_{n=0}^{\infty}\frac{|\beta|^{2n}}{[f(n)!]^{2}n!}\right\} ^{\frac{1}{2}},\label{eq:Norm}\end{equation}
 which is independent of time. Third, we get the reference section,
which is \[
|\chi_{0}(t)\rangle=\frac{1}{N_{1}}e^{-i\chi}e^{-\frac{|\beta|^{2}}{2}}\sum_{n=0}^{\infty}\beta^{n}\frac{1}{f(n)!\sqrt{n!}}e^{-iE_{n}t}|n\rangle,\]
 via substituting Eq. \eqref{eq:TimeEvolutionOfFirstCoherentState}
Eq. \eqref{eq:FirstFraction} and Eq. \eqref{eq:Norm} into Eq. \eqref{eq:ReferenceSection}.
It is in phase with the initial reference section $|\chi_{0}(0)\rangle$,
which is in accordance with Pancharatnam condition \cite{pancharatnam1956connection}.
Fourth, the connection is calculated by use of Eq. \eqref{eq:Norm},
which takes this form\begin{equation}
i\langle\chi_{0}(t)|\frac{d}{dt}|\chi_{0}(t)\rangle=\frac{d\chi}{dt}+\text{\ensuremath{\omega}\ensuremath{\ensuremath{\left\{ \frac{1}{2}+|\beta|^{2}\frac{\sum_{n=0}^{\infty}\frac{|\beta|^{2n}}{[f(n)!]^{2}n!}\frac{1}{[f(n+1)]^{2}}}{\sum_{n=0}^{\infty}\frac{|\beta|^{2n}}{[f(n)!]^{2}n!}}\right\} }}}.\label{eq:ConnectionOfFirstCoherentState}\end{equation}
 Finally, the non-unitary and non-cyclic geometric phase is reached,
i.e. ,\begin{equation}
\gamma(t)=-\arctan\frac{\sum_{n=0}^{\infty}\frac{|\beta|^{2n}}{[f(n)!]^{2}n!}\sin(E_{n}t)}{\sum_{n=0}^{\infty}\frac{|\beta|^{2n}}{[f(n)!]^{2}n!}\cos(E_{n}t)}+\text{\ensuremath{\omega t}\ensuremath{\left\{ \frac{1}{2}+|\beta|^{2}\frac{\sum_{n=0}^{\infty}\frac{|\beta|^{2n}}{[f(n)!]^{2}n!}\frac{1}{[f(n+1)]^{2}}}{\sum_{n=0}^{\infty}\frac{|\beta|^{2n}}{[f(n)!]^{2}n!}}\right\} }\ensuremath{~mod~2\pi}},\label{eq:GeometricPhaseOfTheFirstCoherentState}\end{equation}
 by substituting Eq. \eqref{eq:MinusTotalPhase} and Eq. \eqref{eq:ConnectionOfFirstCoherentState}
into Eq. \eqref{eq:GeometricPhase}. It is regarded as the result
of parallel transport with Pancharatnam connection \cite{pancharatnam1956connection},
which uncover the geometric meaning.

\begin{figure}
\includegraphics{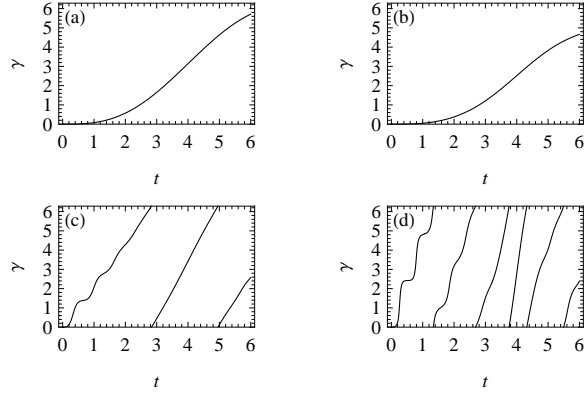}\caption{\label{fig:GeometricPhaseOfTheFirstCoherentState}When we set $|\beta|^{2}=1$
and $\omega=\pi/4$, the geometric phase $\text{\ensuremath{\gamma}}$
with respect to time $t$ according to Eq. \eqref{eq:GeometricPhaseOfTheFirstCoherentState}
together with Eq. \eqref{eq:NonlinearFunction} is shown for various
values of $\eta$. (a), $\eta=0$; (b), $\eta=0.33$; (c), $\eta=0.6$;
(d), $\eta=0.8$.}

\end{figure}

Furthermore, let's discuss about the above geometric phase. Except
for the general properties of geometric phase, such as reparametrization
invariance and manifest gauge invariance under the gauge transformation
$|\chi_{0}^{\prime}(t)\rangle=e^{i2n\pi}|\chi_{0}(t)\rangle$ \cite{pati1995geometric},
where $n$ is integer, let's discuss other ones. To begin with, when
$f(n)=1$, the nonlinear coherent state is reduced to the normal coherent
state, so is the corresponding geometric phase. It takes this form\begin{equation}
\gamma=-\arctan\frac{\sum_{n=0}^{\infty}\frac{|\beta|^{2n}}{n!}\sin(E_{n}t)}{\sum_{n=0}^{\infty}\frac{|\beta|^{2n}}{n!}\cos(E_{n}t)}+\omega t\left(\frac{1}{2}+|\beta|^{2}\right)~mod~2\pi,\label{eq:GeometricPhaseOfNormalCoherentState}\end{equation}
 which coincides with the result \cite{pati1995geometric}. Moreover,
if the quantum state undergoes cyclic evolution, the geometric phase
\eqref{eq:GeometricPhaseOfTheFirstCoherentState} becomes\[
\gamma\left(\frac{2\pi}{\omega}\right)=2\pi|\beta|^{2}\frac{\sum_{n=0}^{\infty}\frac{|\beta|^{2n}}{[f(n)!]^{2}n!}\frac{1}{[f(n+1)]^{2}}}{\sum_{n=0}^{\infty}\frac{|\beta|^{2n}}{[f(n)!]^{2}n!}}~mod~2\pi.\]
 In addition, let's focus on the non-locality of Eq. \eqref{eq:GeometricPhaseOfTheFirstCoherentState}.
Pati \cite{pati1995geometric} has proved that the geometric phase
along the geodesic is equal to zero. Hence, if we join the end points
by a geodesic, the line integral \eqref{eq:GeometricPhase} can be
converted to a surface integral by Stokes's theorem. So the non-locality
for this specific problem is unveiled. Finally, if the different nonlinear
function $f(n)$ is chosen, we will get different geometric phase
\eqref{eq:GeometricPhaseOfTheFirstCoherentState}. In this paper,
we choose \begin{equation}
f(n)=L_{n}^{1}(\eta^{2})[(n+1)L_{n}^{0}(\eta^{2})]^{-1},\label{eq:NonlinearFunction}\end{equation}
 which is important to describe the motion of a ion in a harmonic
potential and interaction with two laser fields \cite{filho1996nonlinear}.
$L_{m}^{n}$ are generalized Laguerre polynomials which are characterized
by the Lamb-Dicke parameter $\eta$. To get some ideas about the evolution
of the geometric phase with respect to time $t$, we show it in Fig.
\eqref{fig:GeometricPhaseOfTheFirstCoherentState}, where we choose
$|\beta|^{2}=1$ and $\omega=\pi/4$. All of these curves vary non-linearly
with time in contrast with dynamical phases, which vary linearly with
time. This is a main difference between the geometric phases and the
dynamical ones. When $\eta=0$, $f(n)=1$, so that the geometric phase
of non-linear coherent states reduce to the standard case \eqref{eq:GeometricPhaseOfNormalCoherentState},
which is shown in Curve $(a)$ in the Fig. \eqref{fig:GeometricPhaseOfTheFirstCoherentState}.
It is illustrated that when $\eta$ increases, the corresponding graph
varies acutely. So we can conclude that non-linear functions tagged
by $\eta$ in this case affect the geometric phase dramatically. The
different functions $f(n)$ in accord with different parameter $\eta$
sped the evolution of the geometric phase up or down, which make the
evolution completely different. So $\eta$ can be regarded as a non-linear
parameter which modulate the evolution of the corresponding geometric
phase. Hence, it probably has potential applications in controlling
or measuring geometric phase. %

\begin{figure}
\includegraphics{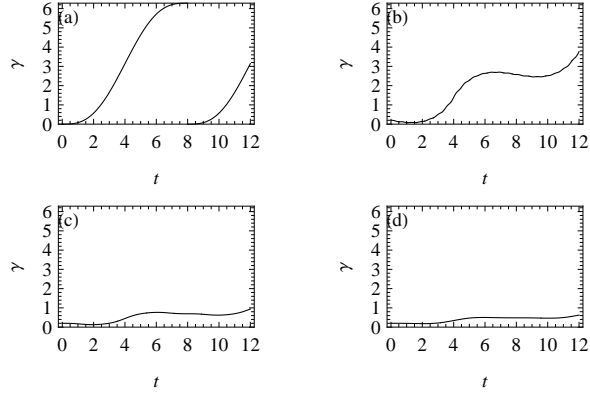}

\caption{\label{fig:GeometricPhaseOfTheSecondCoherentState}When we set $|\beta|^{2}=1$
and $\omega=\pi/4$, the geometric phase $\text{\ensuremath{\gamma}}$
with respect to time $t$ according to Eq. \eqref{eq:GeometricPhaseOfTheSecondCoherentState}
together with Eq. \eqref{eq:NonlinearFunction} is shown for various
values of $\eta$. (a), $\eta=0$; (b), $\eta=0.33$; (c), $\eta=0.6$;
(d), $\eta=0.8$.}

\end{figure}

In the above paragraphs, we have discussed the above non-unitary and
non-cyclic geometric phase of the first nonlinear coherent state.
Now let's study the counterpart of another nonlinear coherent state.
For $t>0$, the time evolution of the state vector reads\[
|\beta(t)\rangle=e^{-\frac{|\beta|^{2}}{2}}\sum_{n=0}^{\infty}\beta^{n}\frac{f(n)!}{\sqrt{n!}}e^{-iE_{n}t}|n\rangle.\]
 By similar calculation, we can obtain the geometric phase, which
is\begin{equation}
\gamma(t)=-\arctan\frac{\sum_{n=0}^{\infty}\frac{|\beta|^{2n}[f(n)!]^{2}}{n!}\sin(E_{n}t)}{\sum_{n=0}^{\infty}\frac{|\beta|^{2n}[f(n)!]^{2}}{n!}\cos(E_{n}t)}+\omega t\left\{ \frac{1}{2}+|\beta|^{2}\frac{\sum_{n=0}^{\infty}\frac{|\beta|^{2n}[f(n)!]^{2}}{n!}[f(n+1)]^{2}}{\sum_{n=0}^{\infty}\frac{|\beta|^{2n}[f(n)!]^{2}}{n!}}\right\} ~mod~2\pi.\label{eq:GeometricPhaseOfTheSecondCoherentState}\end{equation}
 In short, the similar properties with the above case are ignored.
Let's concentrate on the special ones. At first, let's consider the
distinction between the geometric phase of the first nonlinear coherent
state \eqref{eq:GeometricPhaseOfTheFirstCoherentState} and the second
one \eqref{eq:GeometricPhaseOfTheSecondCoherentState}. It is not
very hard to observe that $f(n)$ which is the mark of the non-linearity
is in the denominator position in the first case, whereas it becomes
in the numerator position in the second case. So the geometric phase
plays the pole of a criterion, which tell the geometric difference
of the two coherent states. Moreover, if $f(n)=1$, Eq. \eqref{eq:GeometricPhaseOfTheSecondCoherentState}
is also reduced to the normal case \eqref{eq:GeometricPhaseOfNormalCoherentState}.
In addition, while $|\beta(t)\rangle$ go through cyclic evolution,
the geometric phase \eqref{eq:GeometricPhaseOfTheSecondCoherentState}
is transformed to be\[
\gamma\left(\frac{2\pi}{\omega}\right)=2\pi|\beta|^{2}\frac{\sum_{n=0}^{\infty}\frac{|\beta|^{2n}[f(n)!]^{2}}{n!}[f(n+1)]^{2}}{\sum_{n=0}^{\infty}\frac{|\beta|^{2n}[f(n)!]^{2}}{n!}}~mod~2\pi.\]
Last but not least, again in order to get the ideas of Eq. \eqref{eq:GeometricPhaseOfTheSecondCoherentState},
we show it in Fig. \eqref{fig:GeometricPhaseOfTheSecondCoherentState}.
For the sake of comparison with Fig. \eqref{fig:GeometricPhaseOfTheFirstCoherentState},
we set the parameters $\omega$ and $|\beta|^{2}$ the same values
as the previous case. Again we can see the impact of $f(n)$, which
is shown in Curve $(b)$, $(c)$ and $(d)$. Taking a glance at curves
$(c)$ and $(d)$, we can know there exist non-liner functions that
makes the evolution of the geometric phase extremely slow. So in contrast
with the geometric phase of normal coherent state \eqref{eq:GeometricPhaseOfNormalCoherentState},
it maybe more convenient to measure and control the geometric phase
precisely in experiments. %

\section{geometric phase for nonlinear squeezed state}

\label{sec:squeezed}

In the previous section, the geometric phases of the two non-linear
coherent states are obtained respectively. Now, in this section, we
will research the counterpart of squeezed states. However, the two
non-linear squeezed states coincide with each other. So we can concentrate
on one case, whose state vector of time evolution has this form\begin{equation}
|\zeta(t)\rangle=\mathcal{N}\sum_{n=0}^{\infty}\left(\frac{e^{i\varphi}\tanh(r)}{2}\right)^{n}\frac{f(2n)!\sqrt{(2n)!}}{n!}e^{-iE_{2n}t}|2n\rangle,\label{eq:SqueezedState}\end{equation}
 where $E_{2n}=\omega(\frac{1}{2}+2n).$ Firstly, let's focus on the
first fraction of the reference section \eqref{eq:ReferenceSection},
which is expressed as\begin{equation}
\frac{\langle\zeta(t)|\zeta(0)\rangle}{|\langle\zeta(t)|\zeta(0)\rangle|}=e^{-i\chi}=\exp\left[i\arctan\left(\frac{Im(\langle\zeta(t)|\zeta(0)\rangle)}{Re(\langle\zeta(t)|\zeta(0)\rangle)}\right)\right],\label{eq:FirstFractionOfSqueezedState}\end{equation}
 Where $\chi$ represents the total phase. Hence, we are ready to
calculate the inner product, i.e. ,\begin{equation}
\langle\zeta(t)|\zeta(0)\rangle=|\mathcal{N}|^{2}\sum_{n=0}^{\infty}\left[\frac{\tanh(r)}{2}\right]^{2n}\frac{[f(2n)!]^{2}(2n)!}{(n!)^{2}}\left[\cos(E_{2n}t)+i\sin(E_{2n}t)\right].\label{eq:InnerProductOfSqueezedState}\end{equation}
 By use of Eq. \eqref{eq:FirstFractionOfSqueezedState} and Eq. \eqref{eq:InnerProductOfSqueezedState},
we can obtain\begin{equation}
-\chi=\arctan\frac{\sum_{n=0}^{\infty}\left[\frac{\tanh(r)}{2}\right]^{2n}\frac{[f(2n)!]^{2}(2n)!}{(n!)^{2}}\sin(E_{2n}t)}{\sum_{n=0}^{\infty}\left[\frac{\tanh(r)}{2}\right]^{2n}\frac{[f(2n)!]^{2}(2n)!}{(n!)^{2}}\cos(E_{2n}t)}.\label{eq:MinusTotalPhaseOfSqueezedState}\end{equation}
 Secondly, let's concentrate on the denominator of the second fraction
of the reference section \eqref{eq:ReferenceSection}, i.e., the norm
of $|\zeta(t)\rangle$, which is\begin{equation}
N=\sqrt{\langle\zeta(t)|\zeta(t)}\rangle=\left\{ |\mathcal{N}|^{2}\sum_{n=0}^{\infty}\left[\frac{\tanh(r)}{2}\right]^{2n}\frac{[f(2n)!]^{2}(2n)!}{(n!)^{2}}\right\} ^{\frac{1}{2}}.\label{eq:NormOfSqueezedState}\end{equation}
 Thirdly, substituting Eq. \eqref{eq:SqueezedState}, Eq. \eqref{eq:FirstFractionOfSqueezedState}
and Eq. \eqref{eq:NormOfSqueezedState} into Eq. \eqref{eq:ReferenceSection},
one can get the reference section\begin{equation}
|\chi_{0}(t)\rangle=e^{-i\chi}\frac{\mathcal{N}}{N}\sum_{n=0}^{\infty}\left(\frac{e^{i\varphi}\tanh(r)}{2}\right)^{n}\frac{f(2n)!\sqrt{(2n)!}}{n!}e^{-iE_{2n}t}|2n\rangle,\label{eq:ReferenceSectionOfSqueezedState}\end{equation}
 which is in phase with $|\chi_{0}(t)\rangle$. Fourthly, by use of
Eq. \eqref{eq:ReferenceSectionOfSqueezedState}, we can acquire the
connection\begin{equation}
i\langle\chi_{0}(t)|\frac{d}{dt}|\chi_{0}(t)\rangle=\frac{d\chi}{dt}+\omega\left\{ \frac{1}{2}+\tanh^{2}(r)\frac{\sum_{n=0}^{\infty}\left[\frac{\tanh(r)}{2}\right]^{2n}\frac{[f(2n)!]^{2}(2n)!}{(n!)^{2}}f^{2}(2n+1)f^{2}(2n+2)(2n+1)}{\sum_{n=0}^{\infty}\left[\frac{\tanh(r)}{2}\right]^{2n}\frac{[f(2n)!]^{2}(2n)!}{(n!)^{2}}}\right\} .\label{eq:ConnectionOfSqueezedState}\end{equation}
 Finally, via integration of the above connection according to Eq.
\eqref{eq:GeometricPhase}, the non-unitary and non-cyclic geometric
phase can be achieved, i.e.,\begin{equation}
\gamma(t)=\chi+\omega t\left\{ \frac{1}{2}+\tanh^{2}(r)\frac{\sum_{n=0}^{\infty}\left[\frac{\tanh(r)}{2}\right]^{2n}\frac{[f(2n)!]^{2}(2n+1)!}{(n!)^{2}}f^{2}(2n+1)f^{2}(2n+2)}{\sum_{n=0}^{\infty}\left[\frac{\tanh(r)}{2}\right]^{2n}\frac{[f(2n)!]^{2}(2n)!}{(n!)^{2}}}\right\} ~mod~2\pi.\label{eq:GeometricPhaseOfSqueezedState}\end{equation}
 It is the outcome of parallel transport according to the Pancharatnam
connection \cite{pancharatnam1956connection}.

\begin{figure}
\includegraphics{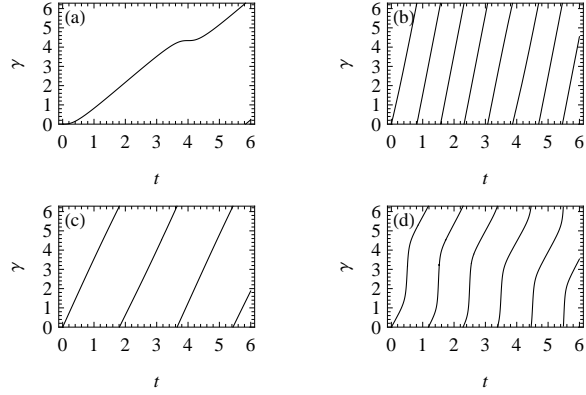}

\caption{\label{fig:GeometricPhaseOfSqueezedState}When we set $r=1$ and $\omega=\pi/4$,
the geometric phase $\text{\ensuremath{\gamma}}$ with respect to
time $t$ according to Eq. \eqref{eq:GeometricPhaseOfSqueezedState}
together with Eq. \eqref{eq:NonlinearFunction} is shown for various
values of $\eta$. (a), $\eta=0$; (b), $\eta=0.0625$; (c), $\eta=0.8$;
(d), $\eta=0.95$.}

\end{figure}

\begin{figure}
\includegraphics{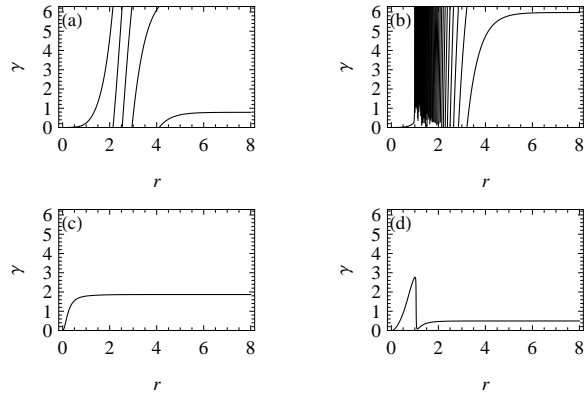}

\caption{\label{fig:SqueezedParameters}When we set $t=0.5$ and $\omega=\pi/4$,
the geometric phase $\text{\ensuremath{\gamma}}$ with respect to
time $r$ according to Eq. \eqref{eq:GeometricPhaseOfSqueezedState}
together with Eq. \eqref{eq:NonlinearFunction} is shown for various
values of $\eta$. (a), $\eta=0$; (b), $\eta=0.0625$; (c), $\eta=0.8$;
(d), $\eta=0.95$.}

\end{figure}

From above calculations, we have obtained the geometric phase \eqref{eq:GeometricPhaseOfSqueezedState},
now let us go on to discuss its features. Above all, when $f(n)=1$,
the above result reduced to the counterpart of normal squeezed state,
which is\[
\gamma(t)=\chi+\omega t\left\{ \frac{1}{2}+\sinh^{2}(r)\right\} ~mod~2\pi.\]
 Moreover, when the quantal state undergoes cyclic evolution, we can
obtain the corresponding geometric phase, which takes this form\[
\gamma(\frac{2\pi}{\omega})=2\pi\tanh^{2}(r)\frac{\sum_{n=0}^{\infty}\left[\frac{\tanh(r)}{2}\right]^{2n}\frac{[f(2n)!]^{2}(2n+1)!}{(n!)^{2}}f^{2}(2n+1)f^{2}(2n+2)}{\sum_{n=0}^{\infty}\left[\frac{\tanh(r)}{2}\right]^{2n}\frac{[f(2n)!]^{2}(2n)!}{(n!)^{2}}}~mod~2\pi.\]
 Again, if $f(n)=1$, the unitary and cyclic geometric phase is gained,
i.e., \[
\gamma(\frac{2\pi}{\omega})=2\pi\sinh^{2}(r)~mod~2\pi.\]
This is the area in phase space enclosed by the phase space trajectory.
In addition, if we join the end-points of the open curve by geodesics,
by use of Stock's theorem, the line integral \eqref{eq:GeometricPhaseOfSqueezedState}
is converted to be surface integral, which unmask the non-locality
of geometric phase of the state. At last, if the same non-linear function
\eqref{eq:NonlinearFunction} are chosen and the parameters are set
as $r=1$ and $\omega=\pi/4$ respectively, we draw a graph of geometric
phase Eq. \eqref{eq:GeometricPhaseOfSqueezedState}, which is shown
in Fig. \eqref{fig:GeometricPhaseOfSqueezedState}. From them, we
can make a conclusion that non-linear functions have a dramatic effect
on the geometric phase. The different functions in accordance with
different $\eta$ make the velocity of evolution of the geometric
phases vary rapidly. On the other hand, it is very interesting to
focus on graphs of $\gamma(r)$ at a given angular velocity $\omega=\pi/4$
and a definite time $t=0.5s$, where $r$ is the squeezed parameter
of Eq. \eqref{eq:GeometricPhaseOfSqueezedState}. The curves are shown
in Fig. \eqref{fig:SqueezedParameters}. It is vividly illustrated
that when $r\rightarrow\infty$, in other words $r$ is sufficiently
large, the corresponding geometric phase may have a plateau, which
is probably a common feature of geometric phase of the squeezed states.
However, the non-linear function tagged by the present parameter $\eta$
affect the limiting value of geometric phase, which is demonstrated
by the curve of (b), (c) and (d) of Fig. \eqref{fig:SqueezedParameters}.%

\section{Conclusion and acknowledgements}

\label{sec:conclusion}

In summary, the non-cyclic and non-unitary geometric phases for nonlinear
coherent and squeezed states are formulated respectively. Furthermore,
some of their properties are discussed such as gauge-invariance, non-locality
and non-linear effects. The non-linear functions have a strong effect
on the evolution of the geometric phase. They speed the evolution
up or down. So this property may have application in control or measure
geometric phase. For the squeezed case, when the squeezed parameter
$r\rightarrow\infty$, the limiting value of the geometric phase is
also modulated by non-linear function at a given time and angular
velocity. In addition, the geometric phases for standard coherent
and squeezed states are recovered if the non-linear function $f(n)=1$.
When the time evolution undergoes a period, their corresponding cyclic
geometric phases are achieved too. 

These authors acknowledge the helpful discussion with Lei Fang and
Ruo-yang zhang. This work was supported in part by NSF of China (Grants
No.10605013 and No.10975075), and the Fundamental Research Funds for
the Central Universities.


\begin{thebibliography}{23}
\bibitem{filho1996nonlinear} de~Matos~Filho~R~L~ and Vogel~W
1996 \newblock {\em Phys. Rev. A} \textbf{54} 4560

\bibitem{roy2000new} Roy~B and Roy~P 2000 \newblock {\em J.
Opt. B: Quantum Semiclass. Opt.} \textbf{2} 65

\bibitem{kwek2003squeezed} Kwek~L~C and Kiang~D 2003 \newblock
{\em J. Opt. B: Quantum Semiclass. Opt.} \textbf{5} 383

\bibitem{berry1984quantal} Berry~M 1984 \newblock {\em Proc.
R. Soc. London Ser. A} \textbf{392} 45

\bibitem{simon1983holonomy} Simon~B 1983 \newblock {\em Phys.
Rev. Lett.} \textbf{51} 2167

\bibitem{aharonov1987phase} Aharonov~Y and Anandan~J 1987 \newblock{\em
Phys. Rev. Lett.} \textbf{58} 1593

\bibitem{samuel1988general} Samuel~J and Bhandari~R 1988 \newblock{\em
Phys. Rev. Lett.} \textbf{60} 2339

\bibitem{pancharatnam1956connection} Pancharatnam~S 1956 \newblock{Proc.
Indian Acad. Sci.{ A}} \textbf{44} 247

\bibitem{pati1995gauge} Pati~A 1995 \newblock {\em J. Phys. A}
\textbf{28} 2087

\bibitem{pati1995geometric} Pati~A 1995 \newblock {\em Phys.
Rev. A} \textbf{52} 2576

\bibitem{manini2000offdiagonal} Manini~N and Pistolesi~F 2000 \newblock
{\em Phys. Rev. Lett.} \textbf{85} 3067

\bibitem{mukunda2001bargmann} Mukunda~N~Arvind Chaturvedi~S and
Simon~R 2001 \newblock {\em Phys. Rev. A} \textbf{65} 012102

\bibitem{kult2007nonabelian} Kult~D 2007 \newblock {\em Europhys.
Lett.} \textbf{78} 60004

\bibitem{uhlmann1986parallel} Uhlmann~A 1986 \newblock {\em Rep.
Math. Phys.} \textbf{24} 229

\bibitem{sjoqvist2000geometric} Sjoquvst~E Pati~A Ekert~A Anandan~J
Ericsson~M Oi~D and Vedral~V 2000 \newblock {\em Phys. Rev.
Lett.} \textbf{85} 2845

\bibitem{tong2003geometric} Singh~K Tong~D Basu~K Chen~J and
Du~J 2003 \newblock {\em Phys. Rev. A} \textbf{67} 032106

\bibitem{tong2004kinematic} Tong~D Sjoqvist~E Kwek~L and Oh C
2004 \newblock {\em Phys. Rev. Lett.} \textbf{93} 080405

\bibitem{shapere1989geometric} Shapere~A and Wilczek~F 1989 \newblock
Geometric phases in physics \newblock World Scientific

\bibitem{bohm2003geometric} Bohm~A Mostafazadeh~A Koizumi~H Niu~Q
and Zwanziger~J 2003 \newblock {\em The geometric phase in quantum
systems} \newblock Springer-Verlag

\bibitem{nayak2008nonabelian} Nayak~C Simon~S Stern~A Freedman~M
and Sarma~S 2008 \newblock {\em Rev. Mod. Phys.} \textbf{80}
1083

\bibitem{abdel2009phase} Bouchene~M~A and Abdel-Aty~M 2009 \newblock{\em
Phys. Rev. A} \textbf{79} 55402

\bibitem{abdel2010sensitivity} Bouchene~M~A, Abdel-Aty~M and Mandal~S
2010 \newblock{\em Phys. Rev. A} \textbf{82} 23409

\bibitem{mukunda1993quantum} Mukunda~N and Simon~R 1993 \newblock
{\em Ann. Phys.} \textbf{228} 205
\end{thebibliography}
\end{document}